\documentclass[%
reprint,
showpacs,preprintnumbers,
amsmath,amssymb,
aps,
prb,
]{revtex4-1}

\usepackage{graphicx}% Include figure files
\usepackage{dcolumn}% Align table columns on decimal point
\usepackage{bm}% bold math
\usepackage{colortbl}
\begin{document}
%-------------------------------------------------

%---------------------------------------------------

\title{Non-stationary spin-filtering effects in correlated quantum dot}

\author{N.\,S.\,Maslova$^{1}$}
\altaffiliation{}
\author{P.\,I.\,Arseyev$^{2,3}$}
\author{V.\,N.\,Mantsevich$^{1}$}
\altaffiliation{} \email{vmantsev@gmail.com}

\affiliation{%
$^{1}$Moscow State University, 119991 Moscow, Russia, $^{2}$ P.N.
Lebedev Physical Institute RAS, 119991 Moscow, Russia,$^3$ Russia
National Research University Higher School of Economics, Moscow,
Russia
}%

\date{\today }
\begin{abstract}
The influence of external magnetic field switching $"$on$"$ and
$"$off$"$ on the non-stationary spin-polarized currents in the
system of correlated single-level quantum dot coupled to
non-magnetic electronic reservoirs has been analyzed. It was shown
that considered system can be used for the effective spin filtering
by analyzing its non-stationary characteristics in particular range
of applied bias voltage.
\end{abstract}

 \maketitle

\section{Introduction}
One of the key issues of spintronics is the control and generation
of spin-polarized currents. Nowadays generation and detection of
spin-polarized currents in semiconductor nanostructures has
attracted great attention since this is the key problem in
developing semiconductor spintronic devices
\cite{Awschalom},\cite{Prinz}, \cite{Crook},\cite{Chuang}. To
generate tunable highly spin-polarized stationary currents the
variety of systems has been already proposed ranging from
semiconductor heterostructures to low-dimensional mesoscopic samples
\cite{Kagan},\cite{Zutic},\cite{Torio},\cite{Shakirov}. Significant
progress has been achieved in experimental and theoretical
investigation of stationary spin-polarized transport in magnetic
tunneling junctions
\cite{Tsymbal},\cite{Zhu},\cite{Ohno},\cite{Fiederling}.
Nevertheless spin-polarized current sources based on the
non-magnetic materials are attractable as one could avoid the
presence of accidental magnetic fields that may result in the
existence of undesirable effects on the spin currents. It was
demonstrated recently that stationary tunneling current could be
spin dependent in the case of non-magnetic leads \cite{Perel'},
\cite{Glazov}. There have been several proposals to generate
stationary spin-polarized currents using non-magnetic materials:
small quantum dots \cite{Recher},\cite{Potok} and coupled quantum
dots \cite{Kagan},\cite{Andrade} built in semiconducting
nanostructures in the presence of external magnetic field. Moreover,
quantum dots systems based on the non-magnetic materials were
proposed as a spin filter prototypes \cite{Koga},
\cite{Voskoboynikov}. Effective spin filtering in such systems
requires to have many quantum dots with the Coulomb correlations
inside each dot \cite{Gong},\cite{Ojeda},\cite{Fu} and also between
the dots \cite{Kagan}.

To the best of our knowledge usually stationary spin-polarized
currents are analyzed. However, creation, diagnostics and
controllable manipulation of charge and spin states in the single
and coupled quantum dots (QDs), applicable for ultra small size
electronic devices design requires analysis of non-stationary
effects and transient processes \cite{Bar-Joseph},\cite{Gurvitz_1},
\cite{Arseyev_1},\cite{Stafford_1},\cite{Hazelzet},\cite{Cota}.
Consequently, non-stationary evolution of initially prepared spin
and charge configurations in correlated quantum dots is of great
interest both from fundamental and technological point of view.

In this paper we analyze non-stationary spin polarized currents
through the correlated single-level QD localized in the tunnel
junction in the presence of applied bias voltage and external
magnetic field, which can be switched $"$on$"$ or $"$off$"$ at a
particular time moment. We demonstrate that single biased QD in the
external magnetic field can be considered as an effective spin
filter based on the analysis of non-stationary spin-polarized
currents, which can flow in the both leads. Currents direction can
be tuned by the external magnetic field switching $"$on$"$ or
$"$off$"$.

\section{Theoretical model}

We consider non-stationary processes in the single-level quantum dot
with Coulomb correlations of localized electrons situated between
two non-magnetic electronic reservoirs  in the presence of external
magnetic field $\textbf{B}$ switched $"$on$"$/$"$off$"$ at $t=t_0$.
The Hamiltonian of the system

\begin{eqnarray}
\hat{H}=\hat{H}_{QD}+\hat{H}_{R}+\hat{H}_{T}
\end{eqnarray}

can be written as a sum of the single-level quantum dot part

\begin{eqnarray}
\hat{H}_{QD}=\sum_{\sigma}\varepsilon_{1}\hat{n}_{1}^{\sigma}++U\hat{n}_{1}^{\sigma}\hat{n}_{1}^{-\sigma},
\end{eqnarray}

non-magnetic electronic reservoirs Hamiltonian

\begin{eqnarray}
\hat{H}_{R}=\sum_{k\sigma}\varepsilon_{k}\hat{c}_{k\sigma}^{+}\hat{c}_{k\sigma}+\sum_{p\sigma}(\varepsilon_{p}-eV)\hat{c}_{p\sigma}^{+}\hat{c}_{p\sigma}
\end{eqnarray}

and the tunneling part

\begin{eqnarray}
\hat{H}_{T}=
\sum_{k\sigma}t_{k}(\hat{c}_{k\sigma}^{+}\hat{c}_{1\sigma}+\hat{c}_{1\sigma}^{+}\hat{c}_{k\sigma})+
\sum_{p\sigma}t_{p}(\hat{c}_{p\sigma}^{+}\hat{c}_{1\sigma}+\hat{c}_{1\sigma}^{+}\hat{c}_{p\sigma}).\nonumber\\
\end{eqnarray}

Here index $k(p)$ labels continuous spectrum states in the leads,
$t_{k(p)}$ is the tunneling transfer amplitude between continuous
spectrum states and quantum dot with the energy level
$\varepsilon_1$ which is considered to be independent on the
momentum and spin. Operators
$\hat{c}_{k(p)\sigma}^{+}/\hat{c}_{k(p)\sigma}$ are the
creation/annihilation operators for the electrons in the continuous
spectrum states $k(p)$.
$\hat{n}_{1}^{\sigma(-\sigma)}=\hat{c}_{1\sigma(-\sigma)}^{+}\hat{c}_{1\sigma(-\sigma)}$-localized
state electron occupation numbers, where operator
$\hat{c}_{1\sigma(-\sigma)}$ destroys electron with spin
$\sigma(-\sigma)$ on the energy level $\varepsilon_1$. $U$ is the
on-site Coulomb repulsion for the double occupation of the quantum
dot. External magnetic field $\textbf{B}$ leads to the Zeeman
splitting of the impurity single level $\varepsilon_1$ proportional
to the atomic $g$ factor. Further analysis deals with the low
temperature regime when the Fermi level is well defined and the
temperature is much lower than all the typical energy scales in the
system. Consequently, the distribution function of electrons in the
leads (band electrons) is close to the Fermi step.

\section{Non-stationary electronic transport formalism}

Let us further consider $\hbar=1$ and $e=1$ elsewhere, so the motion
equations for the electron operators products
$\hat{n}_{1}^{\sigma}$,
$\hat{n}_{1k}^{\sigma}=\hat{c}_{1\sigma}^{+}\hat{c}_{k\sigma}$ and
$\hat{c}_{k^{'}\sigma}^{+}\hat{c}_{k\sigma}$ can be written as:

\begin{eqnarray}
i\frac{\partial \hat{n}_{1}^{\sigma}}{\partial
t}=&-&\sum_{k,\sigma}t_{k}\cdot(\hat{n}_{k1}^{\sigma}-\hat{n}_{1k}^{\sigma}),
\label{1}
\end{eqnarray}

\begin{eqnarray}
i\frac{\partial \hat{n}_{1k}^{\sigma}}{\partial
t}=&-&(\varepsilon_{1}^{\sigma}-\varepsilon_k)\cdot
\hat{n}_{1k}^{\sigma}-U\cdot\hat{n}_{1}^{-\sigma}
\hat{n}_{1k}^{\sigma}+\nonumber\\&+&
t_{k}\cdot(\hat{n}_{1}^{\sigma}-\hat{n}_{k}^{\sigma})
-\sum_{k^{'}\neq
k}t_{k^{'}}\cdot\hat{c}_{k^{'}\sigma}^{+}\hat{c}_{k\sigma},\label{2}
\end{eqnarray}

\begin{eqnarray}
i\frac{\partial \hat{c}_{k^{'}\sigma}^{+}\hat{c}_{k\sigma}}{\partial
t}=&-&(\varepsilon_{k^{'}}-\varepsilon_k)\cdot
\hat{c}_{k^{'}\sigma}^{+}\hat{c}_{k\sigma}-\nonumber\\&-&t_{k^{'}}\cdot
\hat{c}_{1\sigma}^{+}\hat{c}_{k\sigma}+t_{k}\cdot
\hat{c}_{k^{'}\sigma}^{+}\hat{c}_{1\sigma}. \label{3}
\end{eqnarray}

and

\begin{eqnarray}
i\frac{\partial \hat{c}_{p\sigma}^{+}\hat{c}_{k\sigma}}{\partial
t}&=&(\varepsilon_{k}-\varepsilon_p)\cdot
\hat{c}_{p\sigma}^{+}\hat{c}_{k\sigma}+\nonumber\\&+&t_{k}\cdot
\hat{c}_{p\sigma}^{+}\hat{c}_{1\sigma}-t_{p}\cdot
\hat{c}_{1\sigma}^{+}\hat{c}_{k\sigma},
\end{eqnarray}

where $\hat{n}_{k}^{\sigma}=\hat{c}_{k\sigma}^{+}\hat{c}_{k\sigma}$
is an occupation operator for the electrons in the reservoir and
$\varepsilon_{\sigma}=\varepsilon_1+\sigma\mu B$ where
$\sigma=\pm1$. Equations of motion for the electron operators
products $\hat{c}_{1\sigma}^{+}\hat{c}_{p\sigma}$ and
$\hat{c}_{p^{'}\sigma}^{+}\hat{c}_{p\sigma}$ can be obtained from
Eq.(\ref{2}) and Eq.(\ref{3}) correspondingly  by the indexes
substitution $k\leftrightarrow p$ and $k^{'}\leftrightarrow p^{'}$.

Following the logic of Ref. \cite{Maslova} one can get kinetic
equations for the electron occupation numbers operators time
evolution in the case of external magnetic field $\textbf{B}$
switching $"$on$"$ at the time moment $t=t_0>0$:

\begin{eqnarray}
\frac{\partial n_{1}^{\sigma}}{\partial
t}=-2\Theta(t_0-t)\cdot\gamma\times\nonumber\\
\times[n_{1}^{\sigma}-(1-n_{1}^{-\sigma})\cdot
\Phi_{\varepsilon}^{T}(t)-n_{1}^{-\sigma}\cdot
\Phi_{\varepsilon+U}^{T}(t)]-\nonumber\\-
2\Theta(t-t_0)\cdot\gamma\times\nonumber\\
\times[n_{1}^{\sigma}-(1-n_{1}^{-\sigma})\cdot
\Phi_{\varepsilon}^{+T}(t)-n_{1}^{-\sigma}\cdot
\Phi_{\varepsilon+U}^{+T}(t)],\nonumber\\
\frac{\partial n_{1}^{-\sigma}}{\partial
t}=-2\Theta(t_0-t)\cdot\gamma\times\nonumber\\
\times[n_{1}^{-\sigma}-(1-n_{1}^{\sigma})\cdot
\Phi_{\varepsilon}^{T}(t)-n_{1}^{\sigma}\cdot
\Phi_{\varepsilon+U}^{T}(t)]-\nonumber\\-
2\Theta(t-t_0)\cdot\gamma\times\nonumber\\
\times[n_{1}^{-\sigma}-(1-n_{1}^{\sigma})\cdot
\Phi_{\varepsilon}^{-T}(t)-n_{1}^{\sigma}\cdot
\Phi_{\varepsilon+U}^{-T}(t)],\nonumber\\
\label{42}\end{eqnarray}

where $\gamma=\gamma_k+\gamma_p$ and
$\gamma_{k(p)}=\pi\nu_{0}t_{k(p)}^{2}$. $\nu_{0}$ - is the
unperturbed density of states in the leads and

\begin{eqnarray}
\hat{\Phi}_{\varepsilon}^{\pm T}(t)&=&\frac{\gamma_k}{\gamma}\cdot\hat{\Phi}^{\pm}_{k\varepsilon}(t)+\frac{\gamma_p}{\gamma}\cdot\hat{\Phi}^{\pm}_{p\varepsilon}(t),\nonumber\\
\hat{\Phi}_{\varepsilon+U}^{\pm T}(t)&=&\frac{\gamma_k}{\gamma}\cdot\hat{\Phi}^{\pm}_{k\varepsilon+U}(t)+\frac{\gamma_p}{\gamma}\cdot\hat{\Phi}^{\pm}_{p\varepsilon+U}(t),\nonumber\\
\label{401}\end{eqnarray}

where

\begin{eqnarray}
\hat{\Phi}_{\varepsilon}^{\pm}(t)=\frac{1}{2}i\cdot \int
d\varepsilon_{k}\cdot f_{k}^{\sigma}(\varepsilon_{k})\times\nonumber\\\times[\frac{1-e^{i(\varepsilon_1\pm\mu B+i\Gamma-\varepsilon_{k})t}}{\varepsilon_1\pm\mu B+i\Gamma-\varepsilon_{k}}-\frac{1-e^{-i(\varepsilon_1\pm\mu B-i\Gamma-\varepsilon_{k})t}}{\varepsilon_1\pm\mu B-i\Gamma-\varepsilon_{k}}],\nonumber\\
\hat{\Phi}_{\varepsilon+U}^{\pm}(t)=\frac{1}{2}i\cdot \int
d\varepsilon_{k}\cdot f_{k}^{\sigma}(\varepsilon_{k})\times\nonumber\\\times[\frac{1-e^{i(\varepsilon_1\pm\mu B+U+i\Gamma-\varepsilon_{k})t}}{\varepsilon_1\pm\mu B+U+i\Gamma-\varepsilon_{k}}-\frac{1-e^{-i(\varepsilon_1\pm\mu B+U-i\Gamma-\varepsilon_{k})t}}{\varepsilon_1\pm\mu B+U-i\Gamma-\varepsilon_{k}}].\nonumber\\
\label{41}\end{eqnarray}

Initially ($t<t_0$) magnetic field $\textbf{B}$ is absent [$\mu B=0$
in Eqs.(\ref{42})-(\ref{41})] and, consequently, the following
relation is valid
$\hat{\Phi}_{\varepsilon}^{\pm}(t)=\hat{\Phi}_{\varepsilon}(t)$. To
analyze system kinetics in the situation when magnetic field was
initially present in the system and switched $"$off$"$ at $t=t_0$
one can easily generalize Eqs.(\ref{42}) by substitution
$t\leftrightarrow t_0$.

Equations for the localized electrons occupation numbers
$n_{1\pm}^{\sigma}(t)$ can be obtained by averaging Eqs.
(\ref{42})-(\ref{41}) for the operators and by decoupling electrons
occupation numbers in the leads. Such decoupling procedure is
reasonable if one considers that electrons in the macroscopic leads
are in the thermal equilibrium \cite{You,Zheng}. After decoupling
one has to replace electron occupation numbers operators in the
reservoir $\hat{n}_{k}^{\sigma}$ in Eqs. (\ref{42})-(\ref{41}) by
the Fermi distribution functions $f_{k}^{\sigma}$.

\section{Non-stationary spin-polarized currents}

If the initial state is a $"$magnetic$"$ one, non-stationary
spin-polarized currents $I_{k(p)}(t)^{\pm}$ flow in the each contact
lead:

\begin{eqnarray}
I_{k}^{\pm}(t)&=&-2\gamma_{k}\cdot[n_{1}^{\pm\sigma}-(1-n_{1}^{\mp\sigma})\cdot
\Phi_{k\varepsilon}^{\pm}(t)-n_{1}^{\mp\sigma}\cdot
\Phi_{k\varepsilon+U}^{\pm}(t)]-\nonumber\\
&-&2\gamma_{k}\cdot[n_{1}^{\pm\sigma}-(1-n_{1}^{\mp\sigma})\cdot
\Phi_{k\varepsilon}(t)-n_{1}^{\mp\sigma}\cdot \Phi_{k\varepsilon+U}(t)],\nonumber\\
I_{p}^{\pm}(t)&=&-2\gamma_{p}\cdot[n_{1}^{\pm\sigma}-(1-n_{1}^{\mp\sigma})\cdot
\Phi_{p\varepsilon}^{\pm}(t)-n_{1}^{\mp\sigma}\cdot
\Phi_{p\varepsilon+U}^{\pm}(t)]-\nonumber\\
&-&2\gamma_{p}\cdot[n_{1}^{\pm\sigma}-(1-n_{1}^{\mp\sigma})\cdot
\Phi_{p\varepsilon}(t)-n_{1}^{\mp\sigma}\cdot
\Phi_{p\varepsilon+U}(t)],\nonumber\\
\end{eqnarray}

where electron occupation numbers $n_{1}^{\pm\sigma}$ are determined
from the system of Equations (\ref{42}) with the magnetic initial
conditions.

\begin{figure}
\includegraphics[width=90mm]{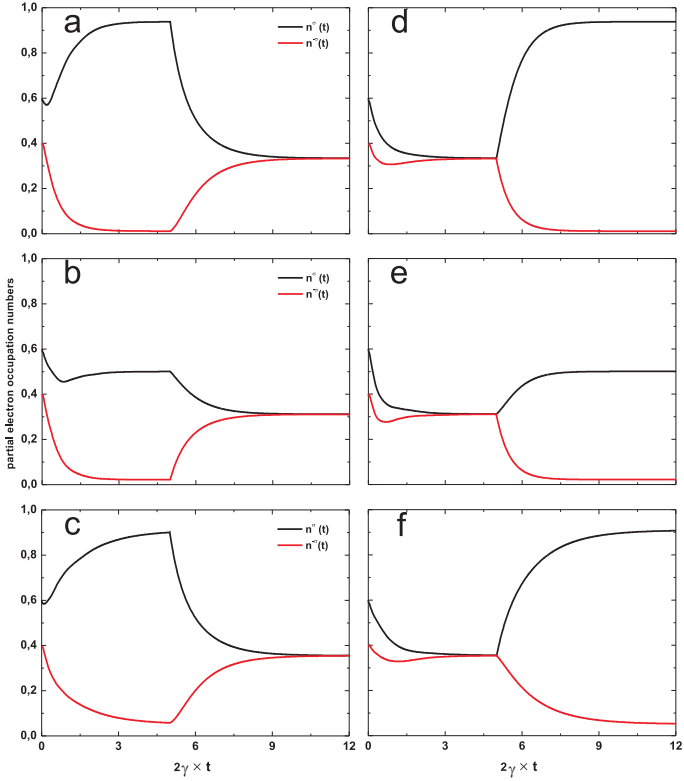}%
\caption{(Color online) Occupation numbers time evolution. Panels
a)-c) correspond to the magnetic field switching $"$on$"$, panels
d)-f)correspond to the magnetic field switching $"$off$"$. a),d)
$\varepsilon_1/2\gamma=-1.25$, $eV/2\gamma=-2.5$; b),e)
$\varepsilon_1/2\gamma=-1.25$, $eV/2\gamma=-7.5$; c),f)
$\varepsilon_1/2\gamma=-5$, $eV/2\gamma=-6.25$. Parameters
$U/2\gamma=10$, $\mu B/2\gamma=-3.25$, $\gamma_k=\gamma_p=\gamma=1$
and initial conditions $n_{1}^{\sigma}(0)=0.6$,
$n_{1}^{-\sigma}(0)=0.4$ are the same for all the figures.}
\label{figure1}
\end{figure}

\begin{figure}
\includegraphics[width=90mm]{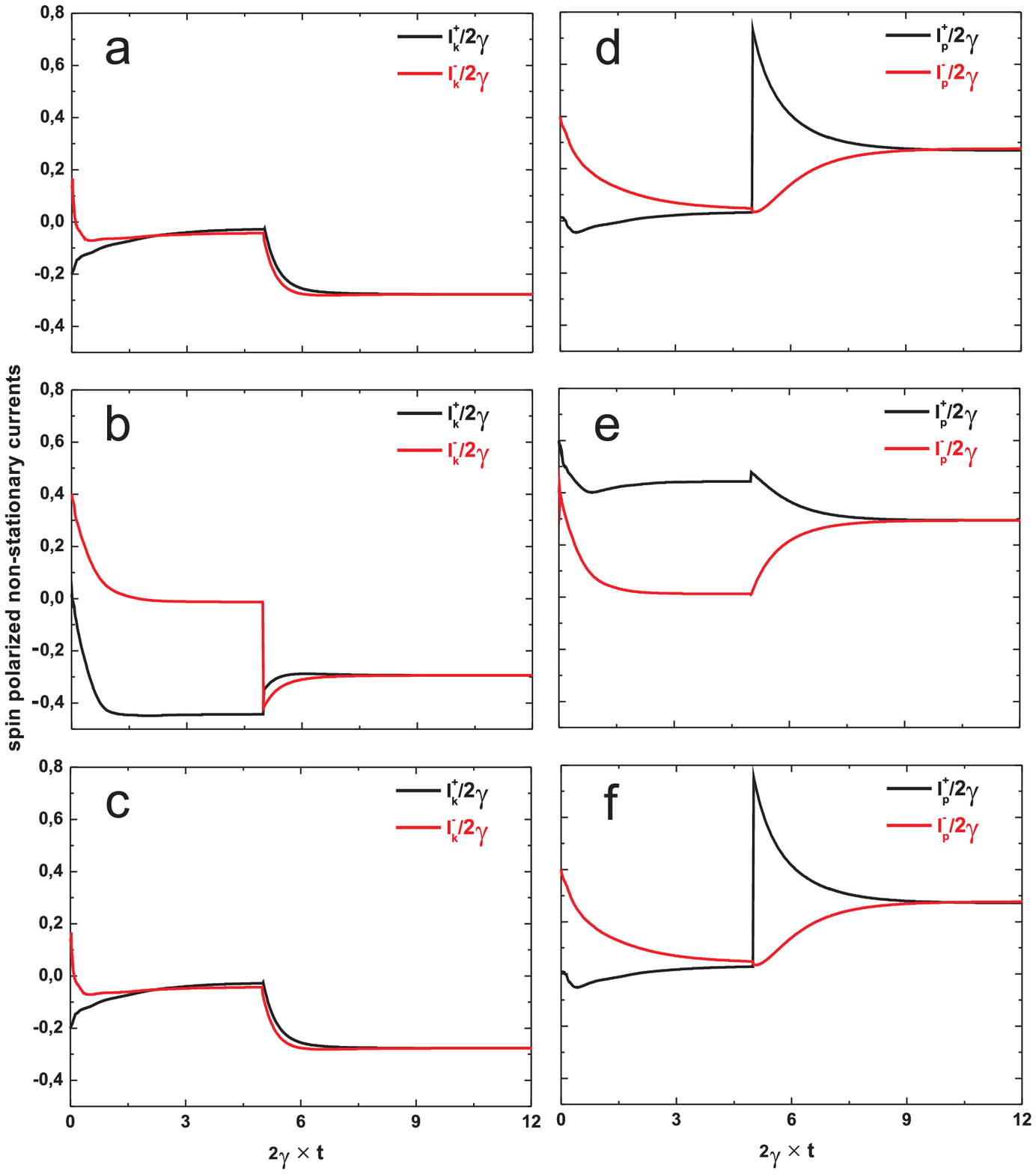}%
\caption{(Color online) Normalized non-stationary spin-polarized
tunneling currents $I_{k(p)^{\pm}(t)}/2\gamma$ in the case of
magnetic field switching $"$off$"$ at $t=t_0$. Panels a)-c)
demonstrate $I_{k^{\pm}(t)}/2\gamma$, panels d)-f)demonstrate
$I_{p^{\pm}(t)}/2\gamma$. a),d) $\varepsilon_1/2\gamma=-1.25$,
$eV/2\gamma=-2.5$; b),e) $\varepsilon_1/2\gamma=-1.25$,
$eV/2\gamma=-7.5$; c),f) $\varepsilon_1/2\gamma=-5$,
$eV/2\gamma=-6.25$. Parameters $U/2\gamma=10$, $\mu
B/2\gamma=-3.25$, $\gamma_k=\gamma_p=\gamma=1$ and initial
conditions $n_{1}^{\sigma}(0)=0.6$, $n_{1}^{-\sigma}(0)=0.4$ are the
same for all the figures.} \label{figure2}
\end{figure}

\begin{figure}
\includegraphics[width=90mm]{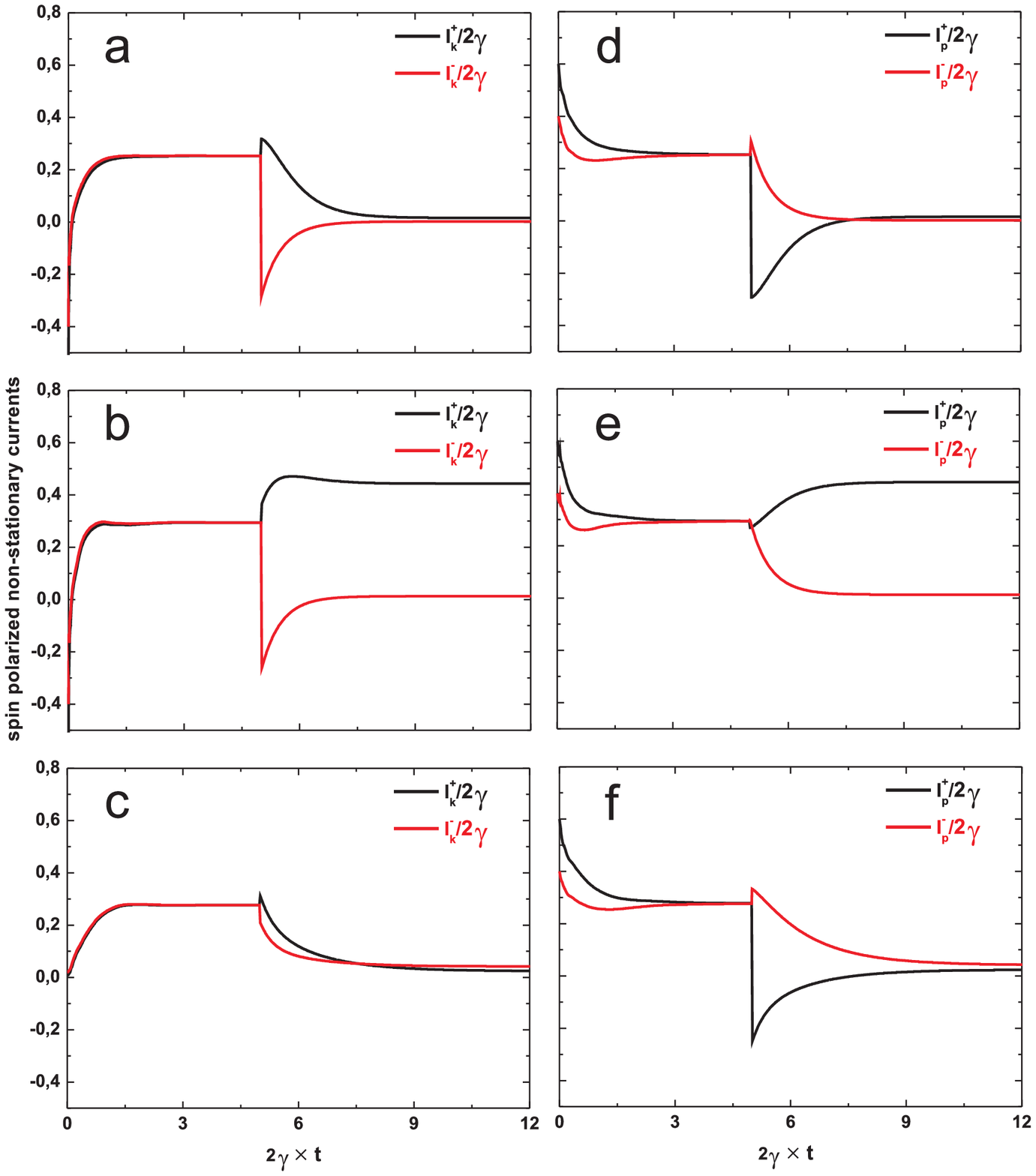}%
\caption{(Color online) Normalized non-stationary spin-polarized
tunneling currents $I_{k(p)^{\pm}(t)}/2\gamma$ in the case of
magnetic field switching $"$on$"$ at $t=t_0$.  Panels a)-c)
demonstrate $I_{k^{\pm}(t)}/2\gamma$, panels d)-f)demonstrate
$I_{p^{\pm}(t)}/2\gamma$. a),d) $\varepsilon_1/2\gamma=-1.25$,
$eV/2\gamma=-2.5$; b),e) $\varepsilon_1/2\gamma=-1.25$,
$eV/2\gamma=-7.5$; c),f) $\varepsilon_1/2\gamma=-5$,
$eV/2\gamma=-6.25$. Parameters $U/2\gamma=10$, $\mu
B/2\gamma=-3.25$, $\gamma_k=\gamma_p=\gamma=1$ and initial
conditions $n_{1}^{\sigma}(0)=0.6$, $n_{1}^{-\sigma}(0)=0.4$ are the
same for all the figures.} \label{figure3}
\end{figure}

\begin{figure}
\includegraphics[width=80mm]{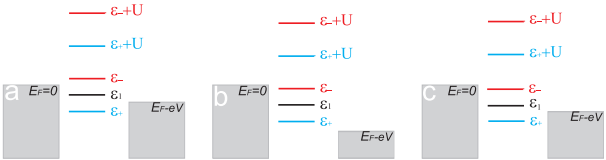}%
\caption{(Color online) Sketch of the correlated QD energy levels
coupled to non-magnetic leads both in the presence ($\varepsilon_+$
and $\varepsilon_-$) and in the absence ($\varepsilon_1$) of
magnetic field.} \label{figure4}
\end{figure}

Non-stationary spin-polarized currents can flow in the both leads
and their direction and polarization can be tuned by magnetic field
$\textbf{B}$ switching $"$on$"$/$"$off$"$. Non-stationary
spin-polarized currents $I_{k(p)}^{\pm}(t)$ behavior for magnetic
field switching $"$on$"$/$"$off$"$ is shown in
Figs.\ref{figure2}-\ref{figure3}. Corresponding electron occupation
numbers behavior is depicted in Fig.\ref{figure1}. Schemes of the QD
energy levels both in the presence ($\varepsilon_+$ and
$\varepsilon_{-}$) and in the absence ($\varepsilon_1$) of magnetic
field are shown in Fig.\ref{figure4}. Let us first focus on the
situation when magnetic field is present at the initial time moment
and switched $"$off$"$ at $t=t_0$ (see Fig.\ref{figure1}a,c,e and
Fig.\ref{figure2}). In the presence of magnetic field when condition
$\varepsilon_{+}<E_F-eV$ occurs (energy level $\varepsilon_{-}$ can
be localized higher or lower than $E_{F}$) (see
Fig.\ref{figure4}a,c), non-stationary spin-polarized currents
$I_{k}^{-}(t)$ and $I_{k}^{+}(t)$ in the lead with $E_F=0$ are
flowing in the same direction (see Fig.\ref{figure2}a,c), contrary
to the currents $I_{p}^{-}(t)$ and $I_{p}^{+}(t)$ flowing in the
opposite directions in the lead with the Fermi level shifted by the
applied bias voltage (lead $p$) (see Fig.\ref{figure2}d,f). In the
stationary state all currents $I_{k(p)}^{\pm}(t)$ values turn to
zero. Magnetic field switching $"$off$"$ results in the appearance
of non-zero spin-polarized currents in both leads. Non-stationary
spin-polarized currents $I_{k}^{-}(t)$ and $I_{k}^{+}(t)$ in the
lead with $E_F=0$ continue flowing in the same direction with the
same non-zero amplitude (see Fig.\ref{figure2}a,c). Currents
$I_{p}^{-}(t)$ and $I_{p}^{+}(t)$ are also flowing in the same
direction but magnetic field switching $"$off$"$ results in the
appearance of total current strong spin polarization at the initial
stage of relaxation as the amplitude of current $I_{p}^{+}(t)$
strongly exceeds the amplitude of non-stationary current
$I_{p}^{-}(t)$ (see Fig.\ref{figure2}d,f). Similar behavior of
electron occupation numbers and non-stationary spin-polarized
currents for two different positions of $\varepsilon_{-}$ (see
Fig.\ref{figure4}a,c) is the result of the Coulomb correlations
presence in the system. In both cases energy level $\varepsilon_{+}$
is occupied and energy level $\varepsilon_{-}$ is unoccupied (even
in the case depicted in Fig.\ref{figure4}c) due to the strong
Coulomb repulsion. Non-stationary current $I_{p}^{+}(t)$ changes
direction with the magnetic field switching $"$off$"$, while
currents $I_{k}^{\pm}(t)$ and $I_{p}^{-}(t)$ are flowing in the same
direction both in presence and in the absence of magnetic field.
This effect can be applied for the effective spin-filtering in the
single QD system alternatively to the previously proposed
spin-filtering mechanisms based on the analysis of multiple QDs
stationary characteristics \cite{Kagan}. Figure\ref{figure4}b
demonstrates that in the presence of magnetic field at the initial
stage of relaxation non-stationary spin-polarized currents
$I_{k}^{\pm}(t)$ can flow in the opposite directions and currents
$I_{p}^{\pm}(t)$ in the same directions (see Fig.\ref{figure2}b,e).
Stationary state reveals the presence of only one non-stationary
current flowing in each lead ($I_{k}^{+}(t)$ and $I_{p}^{+}(t)$
correspondingly). Magnetic field switching $"$off$"$ causes the
appearance of both spin-polarized currents $I_{k}^{\pm}(t)$ and
$I_{p}^{\pm}(t)$ flowing in each lead in the same direction. In the
stationary state spin currents values in each lead become equal.

Electron occupation numbers and non-stationary spin-polarized
currents behavior in the case when magnetic field is absent at the
initial time moment and switched $"$on$"$ at $t=t_0$ is shown in
Fig.\ref{figure1}b,d,f and Fig.\ref{figure3} correspondingly.
Obtained results demonstrate that magnetic field switching $"$on$"$
allows to consider single QD as an effective spin-filter based on
the analysis of its non-stationary characteristics. In the absence
of magnetic field non-stationary spin-polarized currents in each
lead $I_{k}^{\pm}(t)$ and $I_{p}^{\pm}(t)$ are flowing in the same
direction and demonstrate equal non-zero stationary values (see
Fig.\ref{figure3}). Magnetic field switching $"$on$"$ results in the
direction changing of one of the spin-polarized currents in the
leads (see Fig.\ref{figure3}a,d,f). Another possible situation deals
with fast switching $"$off$"$ of one of the spin-polarized currents
in each lead when magnetic field is switched $"$on$"$ ( see
Fig.\ref{figure3}b,e). Consequently, only non-stationary current
with a certain spin orientation continue flowing in each lead in the
presence of magnetic field.

To observe these effects the switching times of magnetic field must
be smaller than the lifetime of the initially prepared magnetic
states. Modern scanning tunneling microscopy/spectroscopy
experiments provide possibility to achieve typical spin-polarized
current values of the order of $10$ pA $\div10$ nA
($1nA\simeq6\times10^{9}e/sec$) (\cite{Amaha},\cite{Fransson}),
which corresponds to the relaxation time scales
$1/\Gamma\simeq1\div100$ nsec for the system parameters depicted in
Fig.\ref{figure2}, Fig.\ref{figure3}.

\section{Conclusion}

We have analyzed the behavior of spin-polarized non-stationary
currents in the system of single-level quantum dot situated between
two non-magnetic electronic reservoirs with Coulomb correlations of
localized electrons in the presence of external magnetic field
switched "on" or "off" at particular time moment. It was
demonstrated that single-level correlated quantum dot can be
considered as an effective spin filter depending on the ration
between the values of magnetic field induced energy level splitting
and applied bias voltage.

This work was supported by RFBR grant $16-32-60024$ $mol-a-dk$ and
by the RF President Grant for young scientists $MD-4550.2016.2$.

 \pagebreak


\begin{thebibliography}{99}

\bibitem{Awschalom}
{\em Semiconductor Spintronics and Quantum Computation}, edited by
D.D. Awschalom, D. Loss, N. Samarth, Nanoscience and Technology
(Springer, Berlin, 2002).
\bibitem{Prinz}
G.A. Prinz, {\em Science} {\bf 282}, 1660, (1998)
\bibitem{Crook}
R. Crook, J. Prance, K.J. Thomas, S.J. Chorley, I. Farrer, D.A.
Ritchie, M. Pepper, C.G. Smith, {\em Science} {\bf 312}, 1359,
(2006)
\bibitem{Chuang}
P. Chuang, S.C. Ho, L.W. Smith, F. Sfigakis, M. Pepper, C.H. Chen,
J.C. Fan, J.P. Griffits, I. Farrer, H.E. Beer, et.al., {\em Nat.
Nanotechnology} {\bf 10}, 35, (2015)
\bibitem{Kagan}
M.Yu. Kagan, V.V. Val'kov, S.V. Aksenov, {\em Phys. Rev. B} {\bf
95}, 035411, (2017)
\bibitem{Zutic}
I. Zutic, J. Fabian, S.D. Sarma, {\em Rev. Mod. Phys.} {\bf 76},
323, (2004)
\bibitem{Torio}
M.E. Torio, K. Hallberg, S. Flach, A.E. Miroshnichenko, M. Titov,
{\em Eur. Phys. J. B} {\bf 37}, 399, (2004)
\bibitem{Shakirov}
A.M. Shakirov, Yu.E. Shchadilova, A.N. Rubtsov, P. Ribeiro, {\em
Phys. Rev. B} {\bf 94}, 224425, (2016)

\bibitem{Tsymbal}
E.Y. Tsymbal, O. Mryasov, P.R. LeClair, {\em J. Phys.: Condens.
Matter} {\bf 15}, R109, (2003)
\bibitem{Zhu}
H.J. Zhu, M. Ramsteiner, H. Kostial, M. Wassermeier, H.-P.
Schonherr, K.H. Ploog, {\em Phys. Rev. Lett.} {\bf 87}, 116601,
(2001)
\bibitem{Ohno}
Y. Ohno, D.K. Young, B. Beschoten, F. Matsukura, H. Ohno, D.D.
Awschalom, {\em Nature}(London) {\bf 402}, 790, (1999)
\bibitem{Fiederling}
R. Fiederling, M. Keim, G. Reuscher, W. Ossau, G. Schmidt, A. Waag,
L.W. Molenkamp, {\em Nature}(London) {\bf 402}, 787, (1999)
\bibitem{Perel'}
V.I. Perel', S.A. Tarasenko, I.N. Yassievich, S.D. Ganichev, V.V.
Bel'kov, W. Prettl, {\em Phys. Rev. B} {\bf 67}, 201304, (2003)
\bibitem{Glazov}
M.M. Glazov, P.S. Alekseev, M.A. Odnoblyudov, V.M. Chistyakov, S.A.
Tarasenko, I.N. Yassievich, {\em Phys. Rev. B} {\bf 71}, 155313,
(2005)
\bibitem{Recher}
P. Recher, E.V. Sukhorukov, D. Loss, {\em Phys. Rev. Lett.} {\bf
85}, 1962, (2000)
\bibitem{Potok}
R.M. Potok, J.A. Folk, C.M. Marcus, V. Umansky, M. Hanson, A.C.
Gossard, {\em Phys. Rev. Lett.} {\bf 91}, 016802, (2003)
\bibitem{Andrade}
J.A. Andrade, P.S. Cornaglia, {\em Phys. Rev. B} {\bf 94}, 235112,
(2016)
\bibitem{Koga}
T. Koga, J. Nitta, H. Takayanagi, S. Datta, {\em Phys. Rev. Lett.}
{\bf 88}, 126601, (2002)
\bibitem{Voskoboynikov}
A. Voskoboynikov, S.S. Liu, C.P. Lee, {\em Phys. Rev. B} {\bf 58},
15397, (1998)
\bibitem{Gong}
W. Gong, Y. Zheng, Y. Liu, T. Lu, {\em Phys. Rev. B} {\bf 73},
245329, (2006)
\bibitem{Ojeda}
J.H. Ojeda, M. Pacheco, P.A. Orellana, {\em Nanotechnology} {\bf
20}, 434013, (2009)
\bibitem{Fu}
H.-H. Fu, K.-L. Yao, {\em Appl. Phys. Lett.} {\bf 100}, 013502,
(2012)
\bibitem{Bar-Joseph}
I. Bar-Joseph, S.A. Gurvitz, {\it Phys.Rev B}, \textbf{44}, 3332,
(1991).
\bibitem{Gurvitz_1}
S.A. Gurvitz, M.S. Marinov, {\it Phys.Rev A}, \textbf{40}, 2166,
(1989).
\bibitem{Arseyev_1}
P.I. Arseyev, N.S. Maslova, V.N. Mantsevich, {\it European Physical
Journal B}, \textbf{85}(7), 249, (2012).
\bibitem{Stafford_1}
C.A. Stafford, N.S. Wingreen, {\it Phys. Rev. Lett.}, \textbf{76},
1916, (1996).
\bibitem{Hazelzet}
B.L. Hazelzet, M.R. Wegewijs, T. H. Stoof, Y.V. Nazarov, {\it Phys.
Rev. B}, \textbf{63}, 165313, (2001).
\bibitem{Cota}
E. Cota, R. Aguado, G. Platero, {\it Phys. Rev. Lett.}, \textbf{94},
107202, (2005).
\bibitem{Maslova}
N.S. Maslova, P.I. Arseyev, V.N. Mantsevich, {\em Solid State Comm.}
{\bf 248}, 21, (2016)
\bibitem{You}
J.Q. You, H.-Z. Zheng, {\em Phys. Rev. B} {\bf 60}, 8727, (1999)
\bibitem{Zheng}
J.Q. You, H.-Z. Zheng, {\em Phys. Rev. B} {\bf 60}, 13314, (1999)
\bibitem{Amaha}
S. Amaha, W. Izumida, T. Hatano, S. Teraoka, S. Tarucha, J. A.
Gupta, and D. G. Austing, {\it Phys. Rev. Lett.}, \textbf{110},
(2013), 016803.
\bibitem{Fransson}
J. Fransson {\it Phys. Rev. B}, \textbf{69}, 201304, (2004).














\end{thebibliography}
\end{document}